\documentclass[aps,prl,twocolumn,showpacs,superscriptaddress,groupedaddress, nofootinbib]{revtex4-1}  
\usepackage{graphicx}  
\usepackage{dcolumn}   
\usepackage{bm}        
\usepackage{amssymb}   
\usepackage{color}
\usepackage[usenames,dvipsnames,svgnames,table]{xcolor}
\definecolor{WHITE}{RGB}{255,255,255}
\usepackage{textcomp, gensymb}
\usepackage{xprintlen}
\usepackage{pgffor}

\begin{document}

\widetext
\title{Mechanical model of Golgi apparatus}
\author{U.~Jeler\v{c}i\v{c}} \affiliation{Jo\v{z}ef Stefan Institute, Jamova 39, SI-1000 Ljubljana, Slovenia}

\begin{abstract}
The Golgi apparatus has intrigued researchers since its discovery and despite the advances, there are still many open questions in regards to its shape and function. Here, we propose a mechanical model of Golgi apparatus stack and explain its most elementary geometrical properties: the equilibrium number of cisternae, the stack size, and its general equilibrium shape. In order to gain comprehensive insight into the stack morphology, we combine both analytical and numerical methods. We successfully reconstruct the stack morphology using minimal assumptions, thus identifying the most dominant factors regulating the stack shape. We demonstrate that energy-wise the stack prefers an overall bent shape and show strong evidence that the adhesion strength determines the equilibrium number of cisternae per stack. We explore the morphological role of fenestrations and discuss their impact on the overall stack structure. We also comment on the effects of the membrane asymmetry on the shape of the cisternae and thus offer a broad steady-state study of the stack morphology and present a method that can be used also for other membrane-bound organelles.
\end{abstract}

\maketitle

\section{Introduction}

The Golgi apparatus is an essential cell organelle found in most eukaryotic cells. It is a highly ordered structure, typically composed of flattened stacks which are surrounded and interconnected by tubular networks and small spherical vesicles \cite{Polishchuk2004, Farquhar1981, Mollenhauer1994, Mollenhauer1991, Ladinsky1999, Ladinsky2002}. Each stack consists of smooth, flattened membrane-bound units referred to as cisternae which are aligned roughly on top of one another. It is typically composed of 4 to 8 cisternae~\cite{Ladinsky1999, Mollenhauer1991} although this number can vary between different species. The number of cisternae per stack is a very well-defined and robust quantity -- it is generally conserved within a specific cell line~\cite{Derganc2006} and is impervious to disassembly and subsequent assembly of the whole organelle~\cite{Cole1996}.

The stack has three basic regions which differ in function, structure, and in microscopic membrane composition: the {\it cis} Golgi, the {\it mid} Golgi, and the {\it trans} Golgi region~\cite{Polishchuk2004, Mogelsvang2004, Rambourg1997}. The {\it cis} side of the organelle is normally convex and represents the part of the stack where matter enters the organelle. The {\it trans} side is on the other side of the stack and is usually concave, and it serves as the exit site for the modified matter. Both {\it cis} and {\it trans} cisternae exhibit buds and dynamic tubules with budding profiles which stretch from their margins~\cite{Ladinsky1999}. Furthermore, a single cisterna is structurally, chemically, and functionally unique~\cite{Mollenhauer1994, Mollenhauer1991}. It is usually large (up to several micrometres in diameter~\cite{Flickinger1969}) and narrow (10 -- 20 nm). The cisternae are stacked close to one another, with typical intercisternal distance of 7 -- 15 nm~\cite{Mollenhauer1991, Flickinger1969} and although the cisternae do not touch physically, they maintain a very uniform distance.

All cisternae are perforated, albeit not homogeneously, with the holes referred to as fenestrations. These can be divided into three classes according to
size~\cite{Ladinsky1999}: i) large well-like openings (\textless~100 nm) are aligned over several consecutive cisternae and often contain budding profiles and free vesicles, ii) medium (65 -- 100 nm) and iii) small holes (\textless~65 nm) are unaligned and are separated from one another by 60 -- 70 nm~\cite{Flickinger1969}. The openings are abundant on the {\it cis} side, decrease in number towards the {\it mid} cisternae and then in general increase again on the {\it trans} side~\cite{Polishchuk2004, Ladinsky1999, Mollenhauer1991} (note that some older works report a different distribution of fenestrations~\cite{Flickinger1969}). Lateral distribution of fenestrations within a single cisterna is also specific with an average cisterna being characterised by a non-fenestrated central region and fenestrated outer parts~\cite{Mollenhauer1991, Flickinger1969}. The shape of the fenestrations is less regular close to the rims of the cisternae and the fenestrations themselves can be, in the extreme case, big enough to create tubules extending from the stack~\cite{Flickinger1969}. 

Even though fenestrations are very abundant, their function is not clear. Ladinsky et al.~\cite{Ladinsky1999} suggest that they act as spot welds to prevent cisternal swelling or they might help in the budding processes by creating regions of high local curvature~\cite{Ladinsky1999, Mogelsvang2004}. The former argument seems to agree with the experimental observations showing a significant difference in the thickness of perforated and unperforated cisternae. The cisternae with fenestrations were found to have a thickness of 50~--~100~nm, whereas the unfenestrated ones had a thickness of only $\approx$~20~nm~\cite{Flickinger1969}. Furthermore, it has been speculated early on that the openings may enable the cisternae to have a greater functional variability due to increased shape pliancy~\cite{Cunningham1966, Tani1975} but the exact connection is yet to be established. 

Current theoretical work and {\it in vitro} experiments are greatly limited to simplified systems of isolated vesicles and vesicle stacks. They are not always directed towards the explanation of the actual organelle shape but rather touch on various phenomena otherwise observed in Golgi apparatus stacks. In the case of unfenestrated structures, most of the research focused on accurately reproducing various stacked erythrocyte morphologies~\cite{Ziherl2007, Svetina2008, Ziherl2008, Skalak1981, Ziherl2007a, Derganc2003} finding that the number of vesicles in the stack depends on the strength of adhesion between them~\cite{Ziherl2007, Svetina2008}. Higher genus vesicles have also been thoroughly investigated (both theoretically and experimentally), albeit only isolated --  within the family of toroidal vesicles~\cite{Zhong-Can1990, Seifert1991, Mutz1991, Fourcade1992} and in the case of vesicles with 2 or 3 holes~\cite{Michalet1995, Julicher1993, Julicher1996}. Further theoretical generalisation to vesicles with an even higher genus is significantly more limited~\cite{Noguchi2015, Akashi2010}. Finally, we point out a recent study by Tachikawa {\it et al.}~\cite{Tachikawa2017} who focused on the morphogenesis of the unfenestrated Golgi apparatus stack and presented an interesting dynamic pathway for the reassembly of the stack structure.

To our knowledge no research successfully addressed highly fenestrated structures in a stacked formation, let alone treated them numerically and with a great degree of generality. The subject of the equilibrium shape of Golgi apparatus stack thus remains open and represents a great challenge both for theoreticians and experimentalists. 

In this paper we address fenestrated Golgi apparatus stacks and present a computational framework (analytical and numerical) based on the Helfrich theory of bending elasticity used to determine their equilibrium shape. Qualitatively, we answer three main questions; what is the equilibrium shape of the stack as predicted by bending energy minimisation; does this shape resemble the native morphology of the Golgi apparatus stack thus justifying the use of elasticity theory as a method of calculating its shape; and is it possible to find a simple mechanism which can determine the equilibrium number of cisternae per stack and their thickness? All in all, we successfully reproduce native-like stack shapes and identify adhesion strength and membrane composition as possible factors determining the stack geometry. We furthermore discuss the role of the fenestrations, membrane asymmetry, and cell environment on the stack shape, thus presenting an exhaustive study of the main aspects of the Golgi apparatus morphology.

\section{Model}

In order to calculate the free energy of Golgi-like vesicle stack the bending energy of each cisterna has to be determined. In our model, each cisterna is approximated as a flattened pancake-like vesicle with volume $V_i$ and surface area $A_i$. The vesicles are then stacked one on top of another and are assumed to adhere {\it via} contact adhesion mechanism. If there are $n$ cisternae in the stack, the total volume and surface area of the stack are then $V=\sum_i^n V_i$ and $A=\sum_i^n A_i$, respectively and the reduced volume of the stack can be defined as $v=6\sqrt{\pi}V/A^{3/2}$.

Since the Golgi stack is observed to preserve its size over a long period of time, the total volume and surface area are assumed to be fixed. However, volumes and surface areas of individual cisternae can vary and thus reflect the dynamic nature of the organelle ({\it i.e.} vesicular transport through the stack and the tubular sections connecting the cistarnae). The free energy of the stack is in our model governed by the bending energy of the cisternal membrane and the adhesion energy due to the association of the cisternae (thermal energy is omitted because we assume that it is suppressed by adhesion). The bending energy of a single cisterna itself is determined using the basic Helfrich bending energy term
\begin{equation}
F_i=2\kappa\int H^2\mbox{d}A+4\pi\kappa_G(1-g), \quad \mbox{for } i=0,...,n
\label{eq:Fi}
\end{equation}
where $H$, $\kappa$, $\kappa_G$ and $g$ denote the mean curvature, mean bending modulus, Gaussian bending modulus and the genus of the vesicle, respectively. The second term in Eq.~(\ref{eq:Fi}), {\it i.e.} the Gaussian term, is a topological invariant and is explicitly taken into account when the cisternae feature highly non-trivial topology ({\it i.e.} many fenestrations) -- it can be omitted when considering structures of the same topological order but should be included in all other general cases, especially if the genus is very high. Finally, the total free energy of the stack is the sum of the bending energies of individual cisternae and the adhesion energy
\begin{equation}
F=\sum_i^n F_i-\Gamma A_c,
\label{eq:F}
\end{equation}
where $\Gamma$ and $A_c$ represent the adhesion strength and the total surface area of the contact zones between vesicles, respectively. The adhesion strength can be alternatively expressed as a reduced quantity $\gamma=\Gamma R_s^2/2\kappa$, where $R_s$ refers to the radius of the sphere with the surface area $A$ and the free energy is measured in the units of $8\pi\kappa$.

We do not include the area difference elasticity term because we assume that the lipid membrane continuously relaxes due to the steady transport along the organelle. The membrane area difference is in our model therefore not fixed albeit non-vanishing -- it amounts to the value at which the total free energy is minimal and differs between stacks with different equilibrium shapes. Furthermore, spontaneous curvature term is also omitted thus neglecting specific composition of the membrane ({\it e.g.} highly conical lipid molecules) and the membrane is assumed to be homogeneous.

In each particular case, the equilibrium configuration is determined by analytically and numerically minimising the free energy [Eq.~(\ref{eq:F})] while keeping the total volume $V$, total surface area $A$ and adhesion strength $\Gamma$ fixed. The number of cisternae in the stack $n$ and the topological order of cisternae are also fixed during an individual minimisation.

\section{Equilibrium shapes of Golgi apparatus stack}

Golgi apparatus stack is topologically extremely complex (involving hundreds of differently sized fenestrations) and thus mathematically very difficult to treat. In general, the analytical models give an excellent insight into the dominant characteristics of the stack, but are limited by the assumptions necessary for the algebra to be manageable. On contrary, numerical solutions offer an exact representation of the actual morphology of the stack within the model, but may be computationally very demanding. A joint approach, used in this paper, overcomes limitations of each individual method and provides us with a comprehensive view of the stack morphology.

\begin{figure}[!ht]
\centering
\includegraphics[width=7cm]{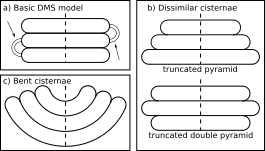}
\caption{Schematic representation of possible stack geometries within the analytical model. a) DMS model where all cisternae are equal and vertically aligned. We show tentative tubular connections (marked with arrows) which represent a possible way of transport of matter, but do not include them in the mechanical model. b) Model with dissimilar cisternae where the radius of the flat part decreases by the same amount in every successive cisterna. c) Model with concentrical curved cisternae forming a nested stack (within the model, the cisternae can bend to hemispherical shape at most). In all cases, we assume that the all the cisternae have the same thickness ({\it i.e.} rim radius) and that the structures are rotationally symmetric with a dashed line marking the symmetry axis (for details see Supplementary Information).}
\label{fig:sch2}
\end{figure}

\textbf{\textit{Analytical results.}} In its most basic form, the Golgi stack can be represented as a system of $n$ identical flat discs as first proposed by Derganc {\it et al.}~\cite{Derganc2006} (Fig.~\ref{fig:sch2}a; termed DMS model). The stack has a fixed total volume and surface area, while the discs themselves can exchange both membrane and lumen. The free energy of such stack combines the bending energy associated with the rims and the adhesion energy (see Supplementary Information). Equilibrium shapes of the stack are determined by minimising the free energy at fixed $V$ and $A$ while varying the adhesion strength $\Gamma$ and the number of cisternae $n$. Derganc {\it et al.} found that at fixed reduced volume $v$ the equilibrium number of cisternae ({\it i.e.} the size of the stack) depends on the reduced adhesion strength. We generalise this result by considering stacks with alternative geometries -- truncated pyramid/truncated double pyramid (where the radial size of each subsequent cisterna differs by the same amount; Fig.~\ref{fig:sch2}b) and curved (with cisternae in the shape of spherical caps; Fig.~\ref{fig:sch2}c).

\begin{figure}[!ht]
\centering
\includegraphics[width=7.5cm]{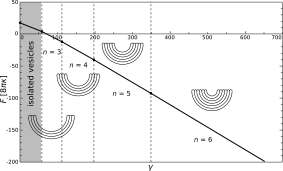}
\caption{Equilibrium solutions within the analytical model. We show free energy as a function of reduced adhesion strength at $v=0.1$. Vertical dashed lines represent the transition boundaries between stacks with different equilibrium number of cisternae, while the insets illustrate the solutions within each region and are drawn to scale. We stress that below the minimal reduced adhesion strength (here $\gamma\approx$ 50) the stack disintegrates into isolated vesicles.}
\label{fig:phase1a}
\end{figure}

When comparing the energies of different stack variants shown in~Fig.~\ref{fig:sch2} (see Supplementary Information), we find that in the regime of relevant reduced adhesion strengths ({\it i.e.} large enough to guarantee that the cisternae form a stack and not a cluster of isolated vesicles) and at small reduced volumes, the DMS model always energy-wise outperforms both pyramidal variants while the curved stack alternative proves to be superior to all three. The actual equilibrium degree of stack curvature depends on several parameters ($n$, $v$, and $\gamma$), but it is always the maximal possible in a given case. Since we artificially limit the maximal curvature, the shapes at biological reduced volumes and within our model conform to the hemispherical shape (Fig.~\ref{fig:phase1a}). We can generalise this result by hypothesising that without the limitation, the stack prefers to be fully curved, {\it i.e.} resembling nested stomatocytes with only a small opening. We therefore analytically show not only that the curved morphology is optimal but also that virtually any equilibrium number of cisternae in the stack is realisable by varying the reduced adhesion strength (Fig.~\ref{fig:phase1a}). Unless too large ($v\gg0.1$), our results qualitatively do not depend on reduced volume.

\textbf{\textit{Numerical results.}} The stack morphology is evaluated numerically using C-based Surface Evolver program package~\cite{brakke}. Within its framework, a surface is represented as a simplicial complex and the equilibrium shape is determined by a gradual minimisation of the predefined free energy -- in our case free energy defined by Eq.~(\ref{eq:F}). As before, we additionally constrain the total volume and total surface area and assume that, in equilibrium, the membrane is relaxed. We search for the deepest local energy minima within the space of physically reasonable shapes set by initial conditions and conjecture the equilibrium morphology by analysing the stack shape and total energy of thus found structures.

\begin{figure}[!ht]
\centering
\includegraphics[width=8.069cm]{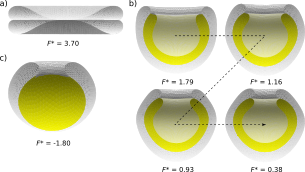}
\caption{Shape of a stack with two cisternae at $v=0.2$ and $\gamma=20$. a) Flat stack (local minimum) and b) bent stack showing several intermediate structures (dashed line indicated the direction of decreasing energy). c) Cup-like stack structure. All the solutions are shown to scale, with adhesion zone marked in yellow, and at the same total volume, surface area, and $\gamma$. $F^*$ represents the free energy in units of $8\pi\kappa$.}
\label{fig:n2v02l0g10b}
\end{figure}

As a proof of principle, and in connection with the analytical results, we first focus on simple systems with a small number of cisternae and start with non-fenestrated stacks. We confirm the analytical prediction that the flat stack does not represent a stable shape and demonstrate that the free energy decreases as the overall bending of the stack increases (Fig.~\ref{fig:n2v02l0g10b}a, b). Apart from being energetically optimal due to the increased contact area, the bent stack shape also features a stable cisternal thicknesses as observed in nature. We note that we find a third group of shapes (besides flat and bent) that has even smaller free energy and resembles a stack of nested stomatocytes where the inner-most cisterna adapts a spherical shape (cup-like structure, Fig.~\ref{fig:n2v02l0g10b}c). We do not include this solution in further analysis because it proves irrelevant ({\it i.e.} geometrically impossible) in the general case of fenestrated stacks and cannot exist at any given small reduced volume. 

\begin{figure}[!ht]
\centering
\includegraphics[width=7.5cm]{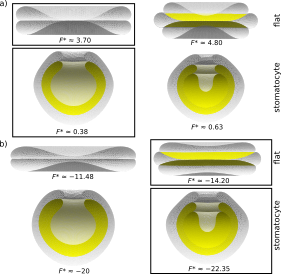}
\caption{Equilibrium number of cisternae in the stack. Flat (local minimum) and nested stomatocyte (global minimum) solution at $v=0.2$ and a) $\gamma=20$ and b) $\gamma=100$ (drawn to scale). In both panels, the equilibrium solution is enclosed in a box and all shapes have the same total volume and surface area. Due to its irrelevance in the general, topologically non-trivial case, we omit showing cup-like solutions.}
\label{fig:v02}
\end{figure}

We extend the analysis to stacks with larger number of cisternae and compute their equilibrium shape at different values of $\gamma$. We find that, in agreement with the analytical solutions, the reduced adhesion strength represents a robust and simple parameter able to determine different equilibrium numbers of cisternae $n_{eq}$ (Fig.~\ref{fig:v02}). Even though we illustrate this behaviour on a simple system and with agreement with the analytical results, we can assume that the trend numerically observed for small $n_{eq}$ continues into the regime of more realistic stack parameters ($n_{eq}\simeq 5$). We thus postulate that a stack with any number of cisternae can be theoretically stabilised by a large enough reduced adhesion strength, although not any value of $\gamma$ may be biologically relevant.

Treating systems of interconnected vesicles with fenestrations is very demanding, albeit unavoidable when discussing more realistic Golgi apparatus models. Apart from purely technical issues connected to the surface triangulation there is also a conceptual problem regarding the accurate description of Gaussian bending energy term. Contrary to the mean bending modulus $\kappa$, the Gaussian bending modulus $\kappa_G$ cannot be easily measured and has been evaluated only for selected systems -- in the case of simple lipid membranes, the majority of experimental and theoretical works find $\kappa_G$ to be negative and close to the value of $\kappa$~\cite{Hu2012, Siegel2004}. On contrary, Golgi apparatus membrane is highly complex and experiments have shown that the regions with high curvature ({\it e.g.} rims and fenestrations) contain specialised proteins~\cite{Day2013} thought to act as shape stabilisers~\cite{Iglic2004, Derganc2007}. The composition of such membrane can thus in principle effect $\kappa_G$, making it either smaller in absolute value or even positive~\cite{Zimmerberg2006}. The sign of $\kappa_G$ in turn strongly effects the energy landscape in heavily fenestrated stacks -- a large negative $\kappa_G$ discourages formation of fenestrations, while positive $\kappa_G$ promotes them. Due to the lack of quantitative estimates, we perform the calculations assuming $\kappa_G=\pm\kappa$ thus illustrating the general impact of fenestrations on stack morphology. We note however, that the value of $\kappa_G$ has no influence on the final shape within an individual minimisation (where $A$, $V$, $\gamma$, and $n$ are fixed). We therefore numerically determine the equilibrium shape by setting $\kappa_G=0$ and consider the Gaussian contribution by adding $4\pi\kappa_G(1-g)$ only when we are interested in the actual value of the free energy ({\it e.g.} when we determine equilibrium number of cistarnae per stack).

\begin{figure}[!ht]
\centering
\includegraphics[width=7.752cm]{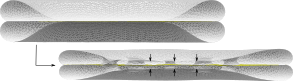}
\caption{Effect of fenestrations on stack morphology -- flat stack at $v=0.12$,  $\gamma=20$ and without (top) or with 10 fenestrations (bottom) per cisterna. Arrows highlight the thickness of the lumen.}
\label{fig:n2v012l10g10FRcros}
\end{figure}

As already mentioned before, the cup-like solution found in the case of trivial topology does not survive the generalisation to a given high genus, because it is physically impossible to have a bulk spherical shape riddled with fenestrations. In the case of flat stacks, we see that the stacks with fenestrations maintain a stable lumen even at very small, biologically relevant reduced volumes, where non-fenestrated stacks collapse (Fig.~\ref{fig:n2v012l10g10FRcros}). This is an important structural feature of the stack consistent with the experimental observations and it is a direct consequence of the introduction of fenestrations. Bent stacks, on the other hand, do not experience such drastic changes in cisternal morphology as flat ones when the genus is increased. At small reduced volumes, the lumen of the cisternae is stable even without fenestrations, although their presence in general guarantees a more uniform thickness along the stack.

Within our approach, the fenestrations are free to move, although they do not move significantly\footnote{This is at least partially contributed to the complexity of the system, where a move by one fenestration effects all the rest through coupling with adhesion surface.}. Initially, we distribute them randomly and in the same fashion in all the cisternae. Based on our solutions it seems that even though the fenestrations adjust their initial positions, they do not tend to cluster, as already established by Michalet {\it et al.}~\cite{Michalet1994} in the case of necks in an infinite membrane sheet. We furthermore find that in cases where the fenestrations are scarce, their radii increase towards the rims of the cisternae which agrees with experimental observations remarkably well. This effect is less prominent in strongly fenestrated stacks where the fenestrations express more uniform radii. The radii themselves are found to depend mostly on the relative ratio between their total number and the reduced volume. As expected energy-wise, the fenestrations are roughly catenoidal in shape and thus contribute very little to the total bending energy of the stack, but due to their specific geometry (neck radius {\it vs.} height ratio) efficiently determine the cisterna thickness. Since the details of fenestrations' sizes and their distribution depend on many parameters, we refrained from further analysis in this direction. It, however, remains a most interesting topic for future work.

\begin{figure}[!ht]
\centering
\includegraphics[width=8.5cm]{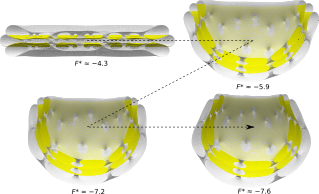}
\caption{Shape of a fenestrated stack with three cisternae at $v=0.09$ and $\gamma=100$. Each cisterna has 30 fenestrations and the solutions are drawn to scale (dashed line shows the direction of decreasing energy) with the same total volume, surface area, and $\gamma$. Flat shape represents the local minimum while the rest illustrate the intermediate structures.}
\label{fig:n3l30g50bending}
\end{figure}

As we consider overall bent stacks, we find a similar trend as in the topologically trivial case -- the free energy of the stack decreases as the bending increases while maintaining a remarkably uniform cisternal thickness at any given overall bending (Fig.~\ref{fig:n3l30g50bending}). The lower energy of curved shapes is also observed in stacks with an even larger number of cisternae, although the numerical analysis is then much more strenuous. It is not entirely clear, however, whether the fenestrated stacks prefer a completely closed stomatocyte morphology seen in stacks without fenestrations, or whether a curved shape with a finite opening is favoured instead. We find that in general, the free energy of a fenestrated stack decreases with the overall curvature of the cisternae at all numbers of cisternae and all numbers of fenestrations per cisterna that we have studied. Based on these results, it is reasonable to expect that the minimal-energy stack is characterised by a small opening (as seen in Fig.~\ref{fig:v02}), but this limit is technically difficult to reach.

\begin{figure}[!ht]
\centering
\includegraphics[width=8.6cm]{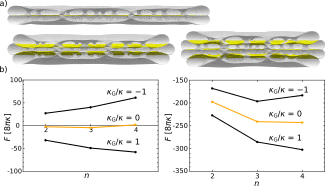}
\caption{Equilibrium number of cisternae in a stack with $v=0.09$ and 30 fenestrations per cisterna. a) Flat stack with different number of cisternae at $\gamma=100$ (drawn to scale). b) Free energy as a function of number of cisternae per stack for $\kappa_G=0$ and $\kappa_G=\pm1$ at $\gamma=100$ (left) and $\gamma=1000$ (right).}
\label{fig:eqnumFen1}
\end{figure}

Finally, we test whether the fenestrated stacks exhibit the same relationship between the equilibrium number of cisternae and reduced adhesion strength that we report in non-fenestrated stacks. Since the curved stomatocyte solutions cannot be easily computed, we perform the analysis using the flat stack solution (Fig.~\ref{fig:eqnumFen1}a) and assume that the results apply also to the general case. We find again that we can qualitatively regulate the equilibrium number of cisternae per stack simply by varying $\gamma$ -- however, the actual value of $n_{eq}$ now depends on the choice of Gaussian bending modulus. For example, if we compare only the bending parts of the free energy ($\kappa_G=0$) we find that in the case of $v=0.09$ and 30 fenestrations per cisterna, the optimal number of cisternae in the stack changes from 3 (at $\gamma=100$) to 4 (at $\gamma=1000$). Taking into account the Gaussian term causes the global energy minimum to shift -- if $\kappa_G<0$ there is an energy penalty per each fenestration and the stack with the smaller number of cisternae (and consequently fenestrations) is preferred. On the other hand, if $\kappa_G>0$ stacks with many cisternae and many fenestrations are stable and as such agree with the native shapes of the organelle (Fig.~\ref{fig:eqnumFen1}b). 

\begin{figure}[!ht]
\centering
\includegraphics[width=8.6cm]{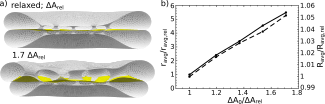}
\caption{Effect of membrane asymmetry on the stack shape at $v=0.2$, $\gamma=20$ and with 10 fenestrations per cisterna. a) Top: stack with relaxed membrane area difference $\Delta A_{rel}$. Bottom: stack where the top cisterna features fixed preferred membrane area difference equal to $1.7\Delta A_{rel}$ and the bottom cisterna is relaxed (drawn to scale). b) The average fenestration radius in top cisterna (left axis, solid line) and the average radius of top cisterna (right axis, dashed line) as functions of the preferred membrane area difference.}
\label{fig:deltaA}
\end{figure}

In order to make our model more realistic, we briefly discuss the effect of membrane asymmetry on the shape of the stack. Even though we still consider that the majority of the stack relaxes the asymmetry {\it via} inter-organelle transport, we assume that outmost {\it cis} cisterna\footnote{We can easily generalise the approach to include other combinations of fixed $\Delta A_0$ or even a gradient along the stack.} maintains a fixed preferred membrane area difference $\Delta A_0$. This simulates the matter entering the organelle {\it via} small spherical vesicles carrying large amounts of $\Delta A_0$ which is absorbed into the stack membrane after the vesicles fuse with the cisterna. A portion of the initial membrane area difference relaxes through membrane rearrangement ({\it e.g.} flip-flop mechanism) and the rest provokes a shape modification of the cisterna. In this case we use a modified bilayer-couple model and find that even a small increase of membrane area difference in comparison to the relaxed value in the rest of the stack noticeably changes the appearance of the affected cisterna (Fig.~\ref{fig:deltaA}a). The radius of the cisterna increases and, even more dramatically, the radii of the fenestrations greatly increase in order to maximise the total surface area of curved regions capable of accommodating incoming $\Delta A_0$ (Fig.~\ref{fig:deltaA}b). We expect a similar behaviour also on the {\it trans} side, although this is more difficult to explain -- contrary to the {\it cis} side where $\Delta A_0$ increases passively via inter-organelle transport, on the {\it trans} side the membrane asymmetry has to be first actively generated in order for the vesiculation and secretion to occur. 

We hypothesise that en even greater increase of  $\Delta A_0$ results in such an enlargement of fenestrations that the cisterna transitions from the flattened perforated shape towards the form of a reticular network. This is in agreement with the observations of the native shape of Golgi apparatus, where outmost cisternae feature large openings and often continue into reticular/tubular networks.

\section{Discussion}

\textbf{\textit{Overall shape of the stack.}} Using simple models, we show that a bent stack (global solution) is in general always energetically preferred over a flat one (local solution) which is consistent with observations and reflects the native morphology of the majority of the mammalian Golgi apparatus stacks. As expected, based on the model assumptions, the exact degree of curvature is not reproduced accurately in our solutions and there are several conceptual improvements that can be implemented easily, albeit are very time-consuming ({\it e.g.} dissimilar $\kappa$ along the stack mimicking chemical uniqueness of cisternae). These can provide us with quantitatively more exact results, but do not introduce qualitatively new findings.

\begin{figure}[!ht]
\centering
\includegraphics[width=5.5cm]{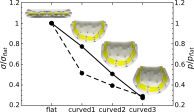}
\caption{Surface tension (left axis, solid line) and pressure (right axis, dashed line) as functions of stack bending. The bending is shown qualitatively and corresponds to shapes from Fig.~\ref{fig:n3l30g50bending} (shown as insets). Both parameters are normalised to their respective values in flat conformation.}
\label{fig:spect}
\end{figure}

Nevertheless, we stress that the whole energy landscape, as presented in this paper, is not necessarily reachable in every biological system. This is consistent with our minimal energy approach and merely takes into account that the solution found in nature is not a global minimum, but a minimum within a set of additional constraints. Together with factors determining the specific composition of membrane, these constraints can be also determined by the biochemical environment ({\it e.g.} $\Delta A_0$, pressure, surface tension, confinement). Surface tension or pressure are especially intriguing because they can be both manipulated {\it via} different cellular processes ({\it e.g.} cytoskeleton stresses or pH) and if we compare their values in stacks with different degrees of bending, significant differences are observed -- compared to the flat solution, surface tension and pressure visibly decrease as the overall curvature of the cisternae increases (Fig.~\ref{fig:spect}).  If, for example, there is a specific range of biological values for surface tension, the equilibrium shape has to be chosen as a minimum energy solution within this range and, depending on the values, less bent stack can be preferred. Given that there are differences in the composition of the Golgi apparatus membrane and the cytosol among species, this argument also offers a possible mechanism for the observed biological variability of stack shapes. 

Apart from general insights into stack morphology, this work presents important new understandings of the role of fenestrations in Golgi apparatus. We see that they are essential for the lumen stability in flat and mildly bend stacks and affect the cisterna morphology in two distinct ways: i) in the central region the fenestrations act as pillars and keep the structure from collapsing, while ii) the fenestrations that are closer to the rims seem to pull the outmost membrane leaflets slightly together and thus have the role of spot welds as hypothesised by Ladinsky {\it et al.}~\cite{Ladinsky1999}. Both mechanisms complement each other in order to create a large region of uniform thickness where the biochemical reactions can take place thus supporting the organelle function. The increase in the cisterna thickness as seen in our analysis is in agreement with the experimental results, where fenestrated cisternae were reported to be up to 5 times thicker than the unfenestrated ones~\cite{Flickinger1969}. Moreover, in a different experiment, temperature blocked cells with an eliminated outward transport were found to have a strongly reduced function and also a significantly decreased number of fenestrations~\cite{Ladinsky1999} -- combining these two observations, we can speculate that the fenestrations ensure well regulated thickness of cisternae, which in turn positively affects the function of the organelle. The size of the fenestrations themselves is found to be transport-mediated and depends on the details of the membrane rearrangement due to the increased membrane area difference in the outer cisternae. Varying only $\Delta A_0$, we can thus control the extent of reticulation, which is consistent with the dynamic nature of the Golgi apparatus peripheral regions.

\vspace{0.2cm}

\textbf{\textit{Equilibrium number of cisternae.}} The relationship between the adhesion strength and equilibrium number of cisternae is found to be universal and is valid regardless of the number of fenestrations or the degree of curvature in the stack. Even though the adhesion strength represents a phenomenological extensive quantity and we cannot determine its value directly from experiments, it is reasonable to expect that it is closely related to the composition of the membrane and of the matrix surrounding the cisternae. Similar is true also for the Gaussian bending modulus of the membrane, which in our model implicitly determines the equilibrium number of cisternae by maximising or minimising the total number of fenestrations. We can assume that the specific protein and lipid configuration of the membrane remains roughly the same within the same cell line but can significantly vary for cells with different roles in the organism (and even more between different species) -- by depending on membrane composition, both parameters allow for great robustness and significant variability. 

\textbf{\textit{Bending elasticity approach.}} Based on our results we conclude that we can use the bending energy minimisation approach to successfully determine the equilibrium shape of a Golgi apparatus stack even by employing the most basic Helfrich model. We show that many of the stack properties derive from a mechanical equilibrium and exhibit a great level of robustness and generality. We thus believe that any other purely function-related processes ({\it e.g.}~\cite{Dmitrieff2013}) that effect the stack morphology serve only to stabilise the shape and consequently have a supporting role in the Golgi apparatus shape determination.

\vspace{0.2cm}

Presented theoretical model introduces a new step in the understanding of the Golgi apparatus shape and shows that significant insights can be gained by using conceptually simple framework and only a few limiting assumptions. Our model thus represents a solid basis for further generalisation and development, especially by taking into account the details of membrane composition ({\it e.g.} variation of $\kappa$ and $\Delta A_0$ along the stack, cytoskeleton, localisation of proteins). The method can be further extended beyond Golgi apparatus morphology and can be applied to other membrane-bound cell organelles, giving a deeper insight into the biomechanics of the cell interior.

\vspace{0.1cm}

{\bf Acknowledgments.} The author thanks P.~Ziherl for his significant contribution to this work and N.~Gov, A.~Bernheim, A.~Lo\v sdorfer-Bo\v zi\v c, G.~Orly, and S.~\v Copar for critically reading the manuscript. This work was supported by the Slovenian Research Agency through Grant No. P1-0055.

\end{document}